\documentclass[aps,prl,floatfix,longbibliography,superscriptaddress,amssymb,amsmath,twocolumn]{revtex4-2}
\pdfoutput=1
\usepackage{graphicx,braket,float}
\usepackage{amsmath}
\usepackage{amsfonts}
\usepackage{amsbsy}
\usepackage{xcolor}
\usepackage{soul}
\usepackage{bm}
\usepackage[normalem]{ulem}
\usepackage[unicode]{hyperref}
\usepackage{braket}
\hypersetup{
    unicode=true, 
    plainpages=false,
    colorlinks=true,
    linkcolor=blue,
    citecolor=blue,
    filecolor=blue,
    urlcolor=blue
}
\begin{document}

\title{Modified Jarzynski equality in a microcanonical ensemble}
\author{L. A. Williamson}
\affiliation{ARC Centre of Excellence for Engineered Quantum Systems, School of Mathematics and Physics, University of Queensland, St Lucia, Queensland 4072, Australia}
\date{\today}

\begin{abstract}
We show that the conventional Jarzynski equality does not hold for a system prepared in a microcanonical ensemble. We derive a modified equality that connects microcanonical work fluctuations to entropy production, in an analogous way to the Jarzynski equality, but with reference to an inverse temperature that depends on the path of the work protocol. For close to isothermal processes the modified equality can improve on the bound $\langle W\rangle\ge \Delta F$. Our result is a special case of a general expression for the microcanonical moment-generating function for any extensive quantity, which enables calculation of the breakdown of ensemble equivalence for thermodynamic fluctuations. We demonstrate our microcanonical Jarzynski equality in a system of driven two-level spins.
\end{abstract}

\maketitle

Ensemble equivalence is a fundamental concept in statistical mechanics. In its simplest form ensemble equivalence argues that microcanonical (fixed energy) and canonical (fixed temperature) averages are equal as long as the mean energies of both ensembles are equal~\cite{gibbs1902}. This equivalence can be derived rigorously for extensive variables when the microcanonical entropy is a concave function of energy~\cite{touchette2015}. For non-extensive variables or fluctuations of extensive variables, however, ensemble equivalence may not hold~\cite{lax1955,lebowitz1967,lanford1973,stroock1991,cancrini2017}.

Fluctuation theorems connect fluctuations of non-equilibrium quantities with equilibrium properties of a system~\cite{sevick2008,campisi2011}. One prominent fluctuation theorem is the Jarzynski equality, which connects non-equilibrium work fluctuations to the work output of the equivalent isothermal process~\cite{jarzynski1997a,jarzynski1997b}. This equality has broad interest not just within physics but also across chemistry and biology, where it aids understanding of chemical and biochemical reactions~\cite{schm2007,rao2018,qian2005,ritort2008} and enables new bioengineering capabilities~\cite{harris2007,gupta2011,berkovich2012}. The Jarzynski equality has been demonstrated in a range of classical and quantum systems, including RNA molecules~\cite{liph2002,coll2005}, mechanical and electronic systems~\cite{doua2005,hoang2018,toya2010,saira2012}, trapped ions~\cite{hube2008,an2015,xiong2018,hahn2023}, ensembles of cold atoms~\cite{ceri2017}, and individual nuclear spins~\cite{liu2023}. Given its broad relevance, it is of interest to explore how well microcanonical-canonical equivalence (MCE) describes the Jarzynski equality, and what modifications are required if MCE breaks down.

The Jarzynski equality is~\cite{jarzynski1997a,jarzynski1997b},
\begin{equation}\label{canJE}
\langle e^{-\beta W}\rangle_\mathrm{c}^\beta=e^{-\beta (F_f(\beta)-F_i(\beta))},
\end{equation}
where $W$ are possible work outputs obtained from a work protocol $H_i\rightarrow H_f$ that need not be adiabatic, with $H_{i(f)}$ the initial(final) Hamiltonian. The notation $\langle \cdot \rangle_\mathrm{c}^\beta$ denotes an average over an initial canonical ensemble at inverse temperature $\beta=(k_B T)^{-1}$. Although fluctuations of $W$ are in general non-equilibrium and path dependent, their combination in Eq.~\eqref{canJE} is path independent and related to the equilibrium free energies $F_{i,f}(\beta)=-\beta^{-1}\ln Z_{i,f}(\beta)$, with $Z_{i,f}(\beta)=\int d\epsilon\, e^{-\beta H_{i,f}(\epsilon)}$ the initial(final) partition function at inverse temperature $\beta$. Here $\int d\epsilon$ denotes an integral over microstates $\epsilon$ of the system with initial(final) energies $H_{i(f)}(\epsilon)$. For quantum systems $\int d\epsilon$ should be replaced by a trace over Hilbert space and the work distribution is sensitive to the act of measurement, however Eq.~\eqref{canJE} remains valid~\cite{tasaki2000,yuka2000,kurchan2000,mukamel2003,chernyak2004,deroeck2004,talkner2007,campisi2010,campisi2011b,hanggi2015,aberg2018,bartolotta2018}.

MCE has been verified for the Jarzynski equality when work is done on a small part of the full system~\cite{park2004,jarzynski2000,cleuren2006,subasi2013} or under constraints on the transition probabilities and density of states~\cite{schmidtke2018,knipschild2021}. However, if the work process drives the system far from equilibrium, MCE is expected to break down~\cite{jeon2015}. Microcanonical forms of Crooks' fluctuation theorem~\cite{crooks1999} have also been derived~\cite{cleuren2006,talkner2008,talkner2013}; however, unlike for canonical ensembles, a general microcanonical Jarzynski equality does not follow straightforwardly from these. Extensions to grand canonical ensembles~\cite{schm2007,yi2011,atas2020} and other generalisations~\cite{gong2015,alhambra2016,hoang2018} have also been derived.

In this paper we show that the Jarzynski equality does not hold in general for a microcanonical initial ensemble. We derive a modified equality in the thermodynamic limit,
\begin{equation}\label{microJE}
\langle e^{-\beta^* W}\rangle_\mathrm{mc}^E=e^{-\beta^*F_f(\beta^*)+\beta F_i(\beta)+(\beta^*-\beta)E},
\end{equation}
where $\langle \cdot\rangle_\mathrm{mc}^E$ is a microcanonical average over initial states with energy $E$ and $\beta$ is the corresponding canonical inverse temperature satisfying $E=\langle H_i\rangle_\mathrm{c}^\beta$. The inverse temperature $\beta^*$ will be given in the main text (Eq.~\eqref{Econ}) and will in general deviate from $\beta$. Importantly, $\beta^*$ depends on $H_i$, $H_f$ and the path of the work protocol connecting these. Hence, unlike Eq.~\eqref{canJE}, the right-hand side of Eq.~\eqref{microJE} is not an equilibrium quantity. For $|\beta^*-\beta|\ll \beta$, Eq.~\eqref{microJE} can be used to improve upon the bound $\langle W\rangle_\mathrm{c}^\beta\ge F_f(\beta)-F_i(\beta)$ (Eq.~\eqref{bound}). Equation~\eqref{microJE} is a special case of a general equation (Eq.~\eqref{momentgeneral}) that quantifies the breakdown of MCE for the moment-generating function for a general extensive variable $X$.

\paragraph{Moment-generating function for work.}
We start by testing MCE for the moment-generating function for work. The moment generating function for an initial microcanonical ensemble is
\begin{equation}\label{microJEstart}
\langle e^{-s W}\rangle_\mathrm{mc}^E=\frac{1}{\Omega(E)}\int d\epsilon\,e^{-s\tilde{H}_f(\epsilon)}e^{sH_i(\epsilon)}\delta(H_i(\epsilon)-E),
\end{equation}
where the $\delta$-function ensures only states within the microcanonical energy shell are included and $\Omega(E)=\int d\epsilon\,\delta (H_i(\epsilon)-E)$ is the multiplicity. The notation $\tilde{H}_f(\epsilon)$ denotes the final energy of the time-evolved microstate $\epsilon$. In the quantum case we assume the work results from a two-point measurement, in which case the integral $\int d\epsilon$ should be replaced by a trace over Hilbert space and $\tilde{H}_f$ should be regarded as a Heisenberg-picture operator $\tilde{H}_f=U^\dagger H_f U$, with $U$ the time evolution operator~\cite{talkner2007}. Note a quantum system in a microcanonical ensemble is diagonal in the energy eigenbasis; hence the initial energy measurement is non-invasive and consideration of initial coherences is not needed (c.f.\ \cite{aberg2018,baumer2018}). Using Eq.~\eqref{microJEstart} the $n$th work moment can be evaluated via $\langle W^n\rangle_\mathrm{mc}^E=(-1)^nd^n\langle e^{-s W}\rangle_\mathrm{mc}^E/ds^n|_{s=0}$.

To test MCE for Eq.~\eqref{microJEstart} we employ the method from~\cite{lax1955} and decompose the microcanonical weighting $\delta(H_i(\epsilon)-E)$ into a sum over (complex) canonical weights,
\begin{equation}\label{decom}
\delta(H_i(\epsilon)-E)=\frac{1}{2\pi i}\int_{-i\infty}^{i\infty} du\,e^{-u(H_i(\epsilon)-E)}.
\end{equation}
Substituting Eq.~\eqref{decom} into Eq.~\eqref{microJEstart} gives~\cite{talkner2008}
\begin{equation}\label{eav}
\langle e^{-s W}\rangle_\mathrm{mc}^E=\frac{1}{\Omega(E)}\frac{1}{2\pi i}\int_{-i\infty}^{i\infty} du\,e^{uE+\ln Q(u,s)},
\end{equation}
where
\begin{equation}\label{Qdef}
Q(u,s)=\int d\epsilon\,e^{-s\tilde{H}_f(\epsilon)}e^{(s-u)H_i(\epsilon)}=Z_i(u)\langle e^{-sW}\rangle_\mathrm{c}^u.
\end{equation}
As a function of $s$ the quantity $Q(u,s)$ interpolates between $Q(u,0)=Z_i(u)$ and $Q(u,u)=Z_f(u)$.

Equation~\eqref{eav} is an exact but in many cases intractable result. In the thermodynamic limit the integral in Eq.~\eqref{eav} can be evaluated using a saddle-point approximation~\cite{lax1955,wong1989}. The saddle point $u=u^*$ is obtained by solving
\begin{equation}\label{saddle}
\frac{\partial}{\partial u^*}\left[u^*E+\ln Q(u^*,s)\right]=0,
\end{equation}
with $u^*$ a function of both $E$ and $s$. Equation~\eqref{saddle} gives $u^*(E,s)$ as the solution to
\begin{equation}\label{saddleSolve}
E=\frac{\langle e^{-s W}H_i\rangle_\mathrm{c}^{u^*}}{\langle e^{-s W}\rangle_\mathrm{c}^{u^*}}.
\end{equation}
In general $u^*$ will differ from the canonical inverse temperature $\beta$ corresponding to $E$ when $s\ne 0$. To implement the saddle-point approximation the path of integration in Eq.~\eqref{eav} is deformed so that it runs through $u^*$ along a small region on the real axis. This manipulation is allowed when the integrand is an entire function of $t$ (i.e.\ holomorphic everywhere in the complex plane)~\cite{ahlfors1979}. In the thermodynamic limit the integrand is dominated by its value at the saddle point, which gives~\cite{wong1989}
\begin{equation}\label{microJE1}
\begin{split}
\langle e^{-s W}\rangle_\mathrm{mc}^E&=\frac{Z_i(u^*)e^{u^*E+O(\ln\lambda)}}{\Omega(E)}\langle e^{-s W}\rangle_\mathrm{c}^{u^*}\\
&=\frac{Z_i(u^*)e^{(u^*-\beta)E+O(\ln\lambda)}}{Z_i(\beta)}\langle e^{-s W}\rangle_\mathrm{c}^{u^*}.
\end{split}
\end{equation}
Corrections beyond the saddle-point approximation are $O(\ln\lambda)$ with $\lambda$ an extensive quantity; these can be neglected in the thermodynamic limit. The second equality in Eq.~\eqref{microJE1} follows from the first by setting $\Omega(E)e^{-\beta E}=Z_i(\beta)$, which is justified by MCE and follows from the above derivation with $s=0$~\cite{lax1955}. As a consistency check Eq.~\eqref{microJE1} can be used to derive Eq.~\eqref{canJE} by integrating over the energy $E$~\footnote{In detail,
\begin{equation*}
\begin{split}
\langle e^{-s W}\rangle_\mathrm{c}^s&=\frac{1}{Z_i(s)}\int dE\,\langle e^{-s W}\rangle_\mathrm{mc}^E\Omega(E)e^{-s E}\\
&=\frac{1}{Z_i(s)}\int dE\,Z_i(u^*)e^{(u^*-s)E}\langle e^{-s W}\rangle_\mathrm{c}^{u^*}\\
&=\frac{1}{Z_i(s)}\int dE\,e^{\ln Q(u^*,s)+(u^*-s)E}.
\end{split}
\end{equation*}
The integral can be evaluated using Laplace's method (the real-space equivalent of the saddle-point approximation), which gives
\begin{equation*}
\frac{d}{dE}\left[\ln Q(u^*,s)+u^*E-s E\right]=0.
\end{equation*}
This combined with Eq.~\eqref{saddle} identifies the saddle-point energy as the energy that gives $u^*=s$. Noting that $Q(s,s)=Z_f(s)$ gives Eq.~\eqref{canJE}.}.

Equation~\eqref{microJE1} together with Eq.~\eqref{saddleSolve} relates the generating function for microcanonical work moments to the canonical moment-generating function at inverse temperature $u^*$. This gives gives $\langle W\rangle_\mathrm{mc}^E=\langle W\rangle_\mathrm{c}^\beta$ but in general gives $\langle W^n\rangle_\mathrm{mc}^E\ne \langle W^n\rangle_\mathrm{c}^\beta$ for $n>1$ (note $du^*/ds\ne 0$ must be taken into account when evaluating moments from Eq.~\eqref{microJE1}). The prefactor in Eq.~\eqref{microJE1} can be written as
\begin{equation}\label{prefactor}
\frac{Z_i(u^*)}{Z_i(\beta)}e^{(u^*-\beta)E}=e^{D(\rho_{H_i}^\beta||\rho_{H_i}^{u^*})},
\end{equation}
with $D(\rho||\sigma)=\int d\epsilon\, \rho(\epsilon)\ln (\rho(\epsilon)/\sigma(\epsilon))$ the relative entropy (Kullback-Leibler divergence) between probability distributions $\rho(\epsilon)$ and $\sigma(\epsilon)$~\cite{cover2006} and $\rho_H^\theta(\epsilon)=e^{-\theta H(\epsilon)}/Z_H(\theta)$ the canonical probability distribution with Hamiltonian $H$ at some inverse temperature $\theta$. (In the quantum case $D(\rho||\sigma)=\operatorname{Tr}(\rho\ln\rho)-\operatorname{Tr}(\rho\ln\sigma)$ for density matrices $\rho$ and $\sigma$~\cite{vedral2002}.) The relative entropy is always non-negative and quantifies the closeness of two distributions with $D(\rho||\sigma)=0$ if and only if $\rho=\sigma$~\cite{cover2006}. It is easy to show that $D(\rho||\rho_H^\theta)$ gives the total entropy production when an ensemble $\rho$ with Hamiltonian $H$ is thermalized with a reservoir at inverse temperature $\theta$~\footnote{In detail,
\begin{equation*}
\begin{split}
D(\rho||\rho_H^\theta)&=\int d\epsilon\,\rho(\epsilon)\ln\rho(\epsilon)-\int d\epsilon\,\rho(\epsilon)\ln \left[e^{-\theta H(\epsilon)}/Z(\theta)\right]\\
&=\int d\epsilon\,\rho(\epsilon)\ln\rho(\epsilon)+\theta\int d\epsilon\,\rho(\epsilon)H(\epsilon)+\ln Z(\theta)\\
&=\int d\epsilon\,\rho(\epsilon)\ln\rho(\epsilon)+\theta\int d\epsilon\,\rho(\epsilon)H(\epsilon)\\
&\phantom{=}-\int d\epsilon\,\rho_H^\theta(\epsilon)\ln\rho_H^\theta(\epsilon)-\theta\int d\epsilon\,\rho_H^\theta(\epsilon)H(\epsilon)\\
&=\delta S-\theta\delta\langle H\rangle,
\end{split}
\end{equation*}
with $Z(\theta)=\int d\epsilon\,e^{-\theta H(\epsilon)}$.}
\begin{equation}\label{Srelinterpret}
D(\rho||\rho_H^\theta)=\delta S-\theta\delta\langle H\rangle,
\end{equation}
with $\delta S$ and $\delta \langle H\rangle$ the change in system entropy (von Neumann entropy in the quantum case) and mean energy, respectively, when $\rho$ is thermalized.

\paragraph{Modified Jarzynski equality.}
Setting $s=\beta$ in Eq.~\eqref{microJE1} gives
\begin{equation}\label{microJEtrial}
\langle e^{-\beta W}\rangle_\mathrm{mc}^E=\frac{Z_i(u^*)e^{(u^*-\beta)E}}{Z_i(\beta)}\langle e^{-\beta W}\rangle_\mathrm{c}^{u^*}
\end{equation}
with $u^*=u^*(E,\beta)$. Comparing with Eq.~\eqref{canJE} shows that MCE will in general break down for the Jarzynski equality when $u^*(E,\beta)\ne \beta$. 

A relation closer in form to Eq.~\eqref{canJE} can be obtained by setting $s=u^*\equiv \beta^*$ in Eq.~\eqref{microJE1} instead. This gives
\begin{equation}\label{microJE2}
\langle e^{-\beta^* W}\rangle_\mathrm{mc}^E=\frac{Z_f(\beta^*)}{Z_i(\beta)}e^{(\beta^*-\beta)E}
\end{equation}
where we have used $\langle e^{-\beta^* W}\rangle_\mathrm{c}^{\beta^*}=Z_f(\beta^*)/Z_i(\beta^*)$ (Eq.~\eqref{canJE}). Equation~\eqref{microJE2}, which is Eq.~\eqref{microJE} in the introduction, is a modified Jarzynski equality for a microcanonical ensemble in the thermodynamic limit and is the main result of this paper. The saddle point $\beta^*$, if it exists, is obtained from Eq.~\eqref{saddleSolve} with $s=\beta^*$,
\begin{equation}\label{Econ}
E=\frac{1}{Z_f(\beta^*)}\int d\epsilon\, H_i(\epsilon)e^{-\beta^*\tilde{H}_f(\epsilon)}.
\end{equation}
The integral in Eq.~\eqref{Econ} can equivalently be replaced by an integral over final microstates due to conservation of phase-space volume. In the quantum case $H_i$ is then a Heisenberg-picture operator time-evolved from $H_f$ according to the reverse work protocol (more formally, $\operatorname{Tr} [H_i e^{-\beta^* U^\dagger H_fU}]=\operatorname{Tr} [UH_iU^\dagger e^{-\beta^* H_f}]$). Hence $\beta^*$ is the inverse temperature of a canonical ensemble $\rho_{H_f}^{\beta^*}$ that is time-evolved to an ensemble $\tilde{\rho}$ with mean energy $E$ under the reverse work protocol. For a cyclic work process Eq.~\eqref{Econ} gives $\beta^*\ge \beta$, since a cyclic work process cannot decrease a thermal ensemble's energy~\cite{pusz1978}. In general the entropy of $\rho_{H_f}^{\beta^*}$ cannot be larger than $\rho_{H_i}^\beta$, since otherwise entropy would decrease upon thermalising $\tilde{\rho}$ to $\rho_{H_i}^\beta$. In the limit of vanishing work, $\tilde{H}_f\rightarrow H_i$, Eq.~\eqref{Econ} gives $\beta^*=\beta$ and MCE is recovered as expected~\cite{park2004,jarzynski2000,cleuren2006,subasi2013,knipschild2021}.

To gain insight into Eq.~\eqref{microJE2} we write it in the form
\begin{equation}\label{interpret1}
\langle e^{-\beta^*(W-\langle W\rangle_\mathrm{mc}^E)}\rangle_\mathrm{mc}^E=e^{D(\rho_f||\rho_{H_f}^{\beta^*})},
\end{equation}
which follows from Eq.~\eqref{Srelinterpret}. Here $\rho_f$ is the ensemble of microstates after implementing the work step. Note $D(\rho_f||\rho_{H_f}^{\beta^*})$ is the same whether $\rho_f$ is computed from a microcanonical or canonical initial ensemble, which follows from Eq.~\eqref{Srelinterpret} and $\langle W\rangle_\mathrm{mc}^E=\langle W\rangle_\mathrm{c}^\beta$. Analogously, Eq.~\eqref{canJE} can be written as~\cite{deffner2010,dorner2012}
\begin{equation}\label{interpret2}
\langle e^{-\beta(W-\langle W\rangle_\mathrm{c}^\beta)}\rangle_\mathrm{c}^\beta=e^{D(\rho_f||\rho_{H_f}^\beta)}.
\end{equation}
Equation~\eqref{interpret1} relates work fluctuations to entropy production $D(\rho_f||\rho_{H_f}^{\beta^*})$, just as for the canonical Jarzynski equality Eq.~\eqref{interpret2}, but with reference to a modified inverse temperature $\beta^*$.

We can use Eq.~\eqref{microJE2} to improve upon the well-known bound $\langle W\rangle_\mathrm{c}^\beta\ge \Delta F$ in cases where $|\beta^*-\beta|\ll\beta$. Expanding Eq.~\eqref{microJE2} in $\delta\beta=\beta^*-\beta$ gives,
\begin{equation}\label{approx}
\frac{1}{\beta^*}\ln\langle e^{-\beta^* W}\rangle_\mathrm{mc}^E=-\Delta F-\Delta S\frac{\delta\beta}{\beta^2}+O(\delta\beta^2),
\end{equation}
where $\Delta F=F_f(\beta)-F_i(\beta)$, $\Delta S=S_f(\beta)-S_i(\beta)$ and $S_{i,f}(\beta)=-\int d\epsilon\,\rho_{H_{i,j}}^\beta\ln\rho_{H_{i,f}}^\beta$. Applying Jensen's inequality~\cite{dekking2005} and $\langle W\rangle_\mathrm{mc}^E=\langle W\rangle_\mathrm{c}^\beta$ to the left-hand side of Eq.~\eqref{approx} then gives,
\begin{equation}\label{bound}
\langle W\rangle_\mathrm{c}^\beta\ge \Delta F+\frac{\Delta S\delta\beta}{\beta^2}+O(\delta\beta^2).
\end{equation}
If $\Delta S\delta\beta>0$ and $|\delta\beta|/\beta\ll 1$, Eq.~\eqref{bound} gives a stronger bound than $\langle W\rangle_\mathrm{c}^\beta\ge \Delta F$, for a given work protocol. The condition $\Delta S\delta\beta>0$ is satisfied when $\Delta S>0$, since $S_f(\beta^*)\le S_i(\beta)<S_f(\beta)$ (see the discussion below Eq.~\eqref{Econ}). Equation~\eqref{bound} then captures the intuitive idea that dissipated work $\langle W\rangle_\mathrm{c}^\beta-\Delta F$ increases with the isothermal entropy change $\Delta S$.

\paragraph{Moment-generating function for a general extensive quantity.}
Equation~\eqref{microJE1} can be generalized to any extensive variable $X$ in place of $W$. Following an identical calculation we obtain
\begin{equation}\label{momentgeneral}
\langle e^{-s X}\rangle_\mathrm{mc}^E=\frac{Z_i(u^*)e^{(u^*-\beta)E}}{Z_i(\beta)}\langle e^{-s X}\rangle_\mathrm{c}^{u^*},
\end{equation}
with $u^*$ obtained from Eq.~\eqref{saddle} with $Q(u,s)=Z_i(u)\langle e^{-sX}\rangle_\mathrm{c}^u$. For a single-time measurement $X$ with Hamiltonian $H$, Eq.~\eqref{momentgeneral} gives
\begin{equation}\label{momentsX}
\begin{split}
\langle X\rangle_\mathrm{mc}^E&=\langle X\rangle_\mathrm{c}^\beta,\\
\langle \delta X^2\rangle_\mathrm{mc}^E&=\langle \delta X^2\rangle_\mathrm{c}^\beta+2\langle \delta X\delta H\rangle_\mathrm{c}^\beta\frac{du^*}{ds}+\langle \delta H^2\rangle_\mathrm{c}^\beta\left(\frac{du^*}{ds}\right)^2\\
&=\langle \delta X^2\rangle_\mathrm{c}^\beta-\frac{(\langle \delta X\delta H\rangle_\mathrm{c}^\beta)^2}{\langle \delta H^2\rangle_\mathrm{c}^\beta},
\end{split}
\end{equation}
with $\delta X=X-\langle X\rangle$. The equality on the last line in Eq.~\eqref{momentsX} follows from differentiating Eq.~\eqref{saddle} with respect to $s$ and solving for $du^*/ds$, and agrees with that derived in~\cite{lebowitz1967,cancrini2017}. Equation~\eqref{momentgeneral} enables a systematic calculation of the breakdown of MCE for thermodynamic fluctuations by taking successive $s$-derivates. This breakdown of MCE is directly related to the relative entropy between the two canonical ensembles $\rho_H^\beta$ and $\rho_H^{u^*}$, see Eq.~\eqref{prefactor}.

\paragraph{Example for a system of two-level spins.}
We demonstrate Eq.~\eqref{microJE2} in a system of $N$ driven two-level spins with Hamiltonian ($\hbar=1$)
\begin{equation}
\hat{H}(t)=\sum_{k=1}^N \left[\omega\hat{\sigma}_+^{(k)}\hat{\sigma}_-^{(k)}+\mathcal{R}(t)(\hat{\sigma}_+^{(k)}+\hat{\sigma}_-^{(k)})\right].
\end{equation}
Here $\hat{\sigma}_+^{(k)}(\hat{\sigma}_-^{(k)})$ denotes the raising(lowering) operator for the $k$th spin with level-spacing $\omega$ and $\mathcal{R}(t)$ is the time-dependent drive. The drive causes transitions between energy levels and does work on the system. For simplicity we suppose $\mathcal{R}$ is a pulse of duration $\tau$ satisfying $\mathcal{R}(0)=\mathcal{R}(\tau)=0$ that is symmetric around $t=\tau/2$~\cite{allen1987}. This pulse causes a spin to flip with probability $p$.

We consider an initial microcanonical ensemble with $M\le N/2$ atoms initially excited, hence $E=M\omega$. The upper bound on $M$ ensures $\beta$ is positive~\cite{ramsey1956}. The microcanonical moment-generating function for work can be evaluated as follows. A work output $W_\ell=\ell\omega-E$, $0\le \ell\le N$, is obtained if $m$ of the initially excited spins remain excited and $\ell-m$ of the initially unexcited spins are excited. This requires $M-m+\ell-m$ spin flips and $N-(M-m+\ell-m)$ spins to remain unchanged. For a fixed microstate this occurs with probability
\begin{equation}\label{pml}
p_{m\rightarrow\ell}=p^{M+\ell-2m}(1-p)^{N-(M+\ell-2m)}.
\end{equation}
The probability of reaching a final state of $\ell$ excited spins from the initial microcanonical ensemble is then
\begin{equation}\label{Pl}
P_\ell=\sum_{m=\operatorname{max}[0,\ell-(N-M)]}^{\operatorname{min}[M,\ell]} \left(\begin{array}{c}M\\m\end{array}\right)\left(\begin{array}{c}N-M\\\ell-m\end{array}\right)p_{m\rightarrow\ell}.
\end{equation}
The limits of the sum ensure $0\le m\le M$ and $0\le \ell-m\le N-M$ and hence all counted quantities are non-negative. The binomial coefficients count the number of ways of choosing the $m$ spins to remain unchanged and the $\ell-m$ spins to flip. Using Eq.~\eqref{Pl} we obtain the microcanonical moment-generating function~\footnote{We write $e^{-s\omega\ell}=e^{-s\omega(\ell-m)-s\omega m}$ and combine the factor $e^{-s\omega(\ell-m)}$ with $p^{\ell-m}$ to obtain,
\begin{equation*}
\begin{split}
\sum_{\ell=m}^{N-M+m}\left(\begin{array}{c}N-M\\\ell-m\end{array}\right)&p^{\ell-m}(1-p)^{N-M-(\ell-m)}e^{-s\omega\ell}\\
&=e^{-s\omega m}(pe^{-s\omega}+1-p)^{N-M},
\end{split}
\end{equation*}
where we have used the binomial identity $\sum_{k=0}^N \left(\begin{array}{c}N\\k\end{array}\right) x^k y^{N-k}=(x+y)^N$. The factor $e^{-s\omega m}$ can be combined with the factor $(1-p)^m$ to obtain Eq.~\eqref{microspin} after using the binomial identity a second time.}
\begin{equation}\label{microspin}
\begin{split}
\langle e^{-s W}\rangle_\mathrm{mc}^E&= e^{sE}\sum_{\ell=0}^N P_\ell e^{-s\omega\ell}\\
&=e^{sE}(1+e^{-s\omega})^N\mathcal{P}(s)^{E/\omega}(1-\mathcal{P}(s))^{N-E/\omega},
\end{split}
\end{equation}
with $\mathcal{P}(s)=(p+(1-p)e^{-s\omega})/(1+e^{-s\omega})$.

\begin{figure}
\includegraphics[trim=0.5cm 2.5cm 1cm 2.2cm,clip=true,width=\columnwidth]{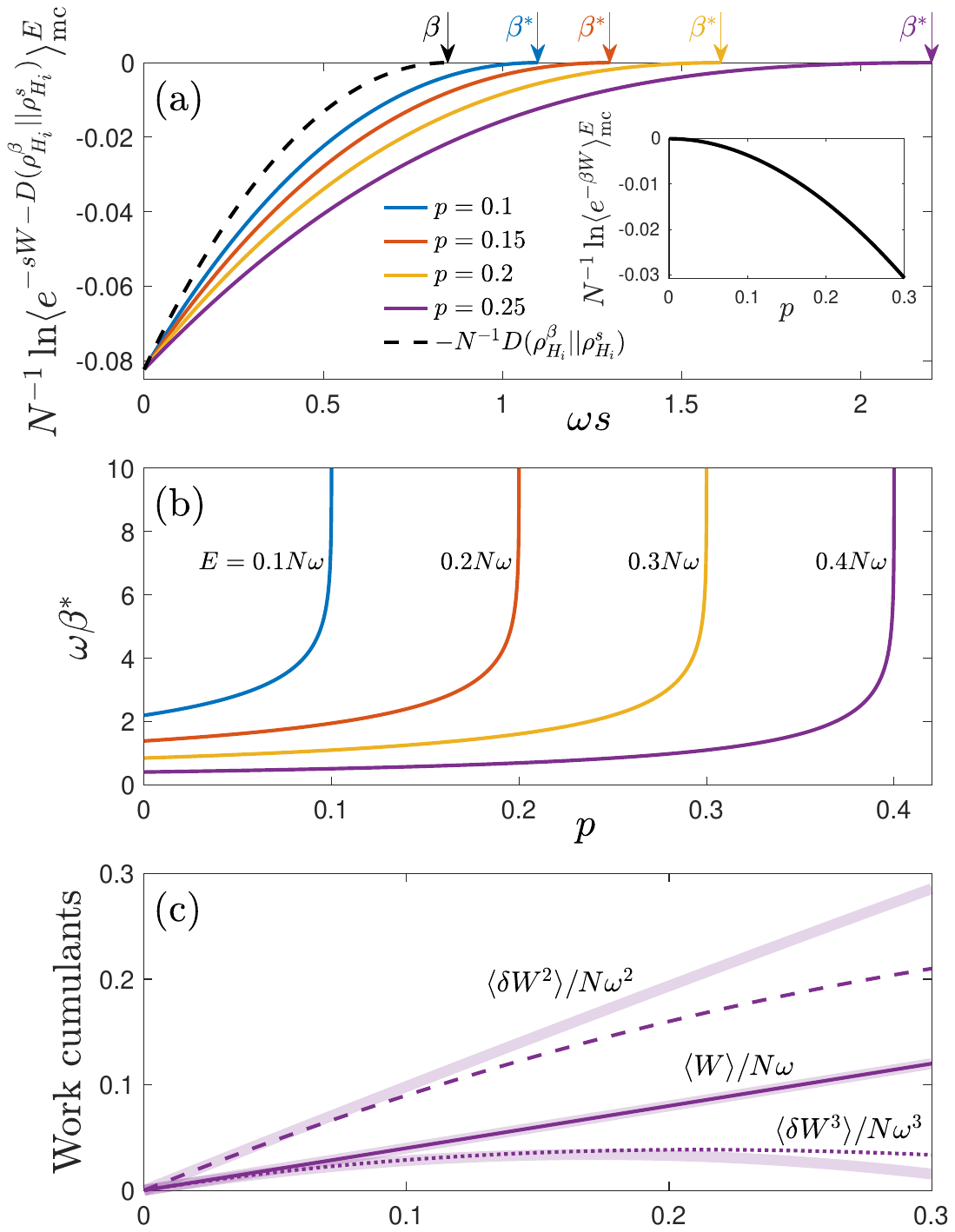}
\caption{\label{fig}(a) Breakdown of the Jarzynski equality for a microcanonical ensemble of cyclically driven ($H_f=H_i$) two-level spins with $E=0.3N\omega$ and four transition rates $p$ (colored lines). Black arrow marks $\beta$ and colored arrows mark $\beta^*$ for each $p$. For a cyclic process Eq.~\eqref{canJE} gives $\ln \langle e^{-\beta W}\rangle_\mathrm{c}^\beta=0$ whereas $\ln\langle e^{-\beta W}\rangle_\mathrm{mc}^E\ne 0$ for $p>0$ (note $D(\rho_{H_i}^\beta||\rho_{H_i}^\beta)=0$). Instead $\ln\langle e^{-\beta^* W-D(\rho_{H_i}^\beta||\rho_{H_i}^{\beta^*})}\rangle_\mathrm{mc}^E=0$ in agreement with Eq.~\eqref{microJE2}. The black-dashed line is $-N^{-1}D(\rho_{H_i}^\beta||\rho_{H_i}^s)$. Inset: $\ln\langle e^{-\beta W}\rangle_\mathrm{mc}^E$ for $E=0.3N\omega$, which deviates from zero for increasing $p$, see Eq.~\eqref{compare}. (b) The inverse temperature $\beta^*$ for four values of $E$. We have $\beta^*>\beta$ for $p>0$ and $\beta^*\rightarrow \infty$ for $p\rightarrow E/N\omega$. (c) First three microcanonical work cumulants $\langle W\rangle_\mathrm{mc}^E$ (dark-solid line), $\langle \delta W^2\rangle_\mathrm{mc}^E$ (dashed line) and $\langle \delta W^3\rangle_\mathrm{mc}^E$ (dotted line) for $E=0.3N\omega$. Corresponding canonical cumulants are shown by faint, thick, solid lines. The $n$th work cumulant has been scaled by $N\omega^n$.}
\end{figure}

To verify Eq.~\eqref{microJE2} we need to obtain $\beta^*$ from Eq.~\eqref{Econ},
\begin{equation}\label{Econspin}
E=N\omega\mathcal{P}(\beta^*).
\end{equation}
We have used that $\mathcal{P}(s)$ is the probability of finding a spin in its excited state after the work process starting from an ensemble $\rho_{H_i}^s$, and noted that the forward and reverse work protocols are identical due to the chosen symmetry of the drive. Equation~\eqref{Econspin} combined with $E=-d\ln Z_i(\beta)/d\beta$ gives $e^{-\beta\omega}=\mathcal{P}(\beta^*)/(1-\mathcal{P}(\beta^*))$. Substituting this into Eq.~\eqref{microspin} with $s=\beta^*$ gives agreement with Eq.~\eqref{microJE2}, see Fig.~\ref{fig}(a). In this example the exact evaluation of the integral Eq.~\eqref{microJEstart} (modified to account for the discreteness of the system) gives the same result as the saddle-point approximation~Eq.~\eqref{microJE1} when $s=\beta^*$. The thermodynamic limit is still invoked, however, by equating $Z_i(\beta)$ with $\Omega(E)e^{-\beta E}$.

The temperature $\beta^*$ obtained from Eq.~\eqref{Econspin} is plotted in Fig.~\ref{fig}(b). For $p=0$ we have $\beta^*=\beta$. Transitions $p>0$ effectively ``heat'' the system~\cite{allahverdyan2005b,polkovnikov2008,polkovnikov2011,santos2011} and hence to reach an ensemble with energy $E$ requires a cooler initial ensemble $\beta^*>\beta$ (see discussion below Eq.~\eqref{Econ}). When $N\omega p=E$, an ensemble with energy $E$ is reached from a zero temperature ($\beta^*=\infty$) initial state; for $N\omega p>E$ Eq.~\eqref{Econspin} has no physical solution.

We quantify deviation from MCE for the Jarzynski equality by computing how far $\ln\langle e^{-\beta W}\rangle_\mathrm{mc}^E$ deviates from its canonical average, which is zero for a cyclic process (see Eq.~\eqref{canJE}). Using Eq.~\eqref{microspin} we obtain
\begin{equation}\label{compare}
\ln\langle e^{-\beta W}\rangle_\mathrm{mc}^E=-2N\sinh^2\left(\frac{\beta\omega}{2}\right)p^2+O(p^3).
\end{equation}
Deviation from MCE therefore grows with the transition rate $p$, see inset to Fig.~\ref{fig}(a), consistent with the increase in $\beta^*-\beta$ (Fig.~\ref{fig}(b)). The deviation from MCE also increases with decreasing temperature as $\ln\langle e^{-\beta W}\rangle_\mathrm{mc}^E=-Np^2\omega^2\beta^2/2+O(\beta^4)$. The first three work cumulants are easily computed from Eq.~\eqref{microspin} and a similar expression for the canonical ensemble, and are plotted in Fig.~\ref{fig}(c)~\footnote{Explicitly,
\begin{equation*}
\begin{split}
\langle W\rangle_\mathrm{mc}^E&=\langle W\rangle_\mathrm{c}^\beta=(N\omega-E)p+E(1-p)-E,\\
\langle \delta W^2\rangle_\mathrm{mc}^E&=N\omega^2 p(1-p)\\
&=\langle \delta W^2\rangle_\mathrm{c}^\beta-4N^{-1}(N\omega-E)E p^2,\\
\langle \delta W^3\rangle_\mathrm{mc}^E&=\omega^2(N\omega-2E)(1-p)(1-2p)p\\
&=\langle \delta W^3\rangle_\mathrm{c}^\beta+8N^{-2}E(N\omega-E)(N\omega-2E)p^3.
\end{split}
\end{equation*}
Note the scaling $\langle \delta W^n\rangle_\mathrm{mc}^E-\langle \delta W^n\rangle_\mathrm{c}^\beta=O(p^n)$ does not hold in general for $n>3$.}. These demonstrate that MCE breaks down beyond the first moment of work, as expected from Eq.~\eqref{momentsX}.

\paragraph{Discussion.}
We have shown that ensemble equivalence does not hold for the Jarzynski equality. We derive a modified relation that connects microcanonical work fluctuations to entropy production in an analogous way to the Jarzynski equality but with reference to a protocol-dependent inverse temperature $\beta^*$. The relation provides a new bound on dissipated work for processes close to isothermal when the corresponding isothermal entropy change is positive. This modified Jarzynski equality is a special case of a more general equation for the microcanonical moment-generating function for any extensive variable. We demonstrate the microcanonical Jarzynski equality in a system of non-interacting two-level spins. Interactions will increase the possible transitions between energy levels, in which case we expect deviation from MCE to be larger.

Our results may be useful when considering fluctuations in quantum systems in pure, or close-to-pure, energy eigenstates, for example in quantum information systems~\cite{auffeves2022}, quantum simulators~\cite{georgescu2014,altman2021} and ultracold-atom experiments. In particular, it would be interesting to consider a microcanonical energy window with some finite width and explore the effect of narrowing this to a single quantum state. The Jarzynski equality has been tested in a system of interacting quantum spins in the context of the eigenstate thermalization hypothesis~\cite{deutsch1991,srednicki1994,rigol2008}, where deviations from a canonical ensemble were identified~\cite{jin2016}. This difference could possibly be accounted for using the microcanonical Jarzynski equality derived here. Generalising our results to include modelling of the work reservoir provides another interesting direction for exploration~\cite{aberg2018}.

\paragraph{Acknowledgements.}
The author thanks Kavan Modi, Thom\'{a}s Fogarty, Thomas Busch and Vishnu Muraleedharan Sajitha for valuable discussions, Andrew Groszek, Karen Kheruntsyan and Blair Blakie for comments on the manuscript, and an anonymous referee for comments that led to Eq.~\eqref{bound}. This research was supported by the Australian Research Council Centre of Excellence for Engineered Quantum Systems (EQUS, CE170100009) and the Australian government Department of Industry, Science, and Resources via the Australia-India Strategic Research Fund (AIRXIV000025).

\end{document}